\documentclass[twocolumn,showpacs,amsmath,amssymb]{revtex4}

\usepackage{graphicx}
\usepackage{dcolumn}
\usepackage{bm}
\newcommand{\etal}{\emph{et al.}}
\newcommand{\be}{\begin{equation}}
\newcommand{\ee}{\end{equation}}
\newcommand{\bfig}{\begin{figure}}
\newcommand{\efig}{\end{figure}}
\newcommand{\incl}{\includegraphics}

\begin{document}      

\title{A hidden constant in the anomalous Hall effect of a high-purity magnet MnSi.
} 
\author{Minhyea Lee$^1$, Y. Onose$^1$, Y. Tokura$^{2,3}$ and N. P. Ong$^1$
}
\affiliation{
$^1$Department of Physics, Princeton University, Princeton, NJ 08544, USA\\
$^2$Department of Applied Physics, University of Tokyo, Tokyo 113-8656, Japan\\
$^3$ERATO, JST, Spin Superstruture Project (SSS), Tsukuba 305-8562, Japan
}

\date{\today}      
\begin{abstract}
Measurements of the Hall conductivity in MnSi can provide incisive tests 
of theories of the anomalous Hall (AH) effect, because both the mean-free-path 
and magnetoresistance (MR) are unusually large for a ferromagnet.  The large MR provides
an accurate way to separate the AH conductivity $\sigma_{xy}^A$ from the ordinary Hall conductivity $\sigma_{xy}^N$.
Below the Curie temperature $T_C$, $\sigma_{xy}^A$ is linearly proportional to $ M$ (magnetization) with a proportionality constant $S_H$ that is independent of both $T$ and $H$.  In particular,  $S_H$ remains a constant while $\sigma_{xy}^N$ changes by a factor of 100 between 5 K and $T_C$.  We discuss implications of the hidden constancy in $S_H$. 
 \end{abstract}

\maketitle                   

The origin of the anomalous Hall (AH) effect in ferromagnets has been vigorously debated for the past 50 years.  Karplus  
and Luttinger (KL)~\cite{Karplus} proposed in 1954 that the AH current is an intrinsic
current that is independent of the mean-free-path $\ell$~\cite{Niu,Nagaosa,JungwirthYao,Haldane}.
In the competing skew-scattering theory, the AH current arises  
from asymmetric scattering off impurities and defects, and is proportional to $\ell$~\cite{Smit}.  
While older experiments favor skew scattering, support for the KL/Berry-phase theory 
has been obtained from recent experiments~\cite{Matl,Tokura,Mathieu04,WLee,Zeng06,Onose}.  
However, uncertainty remains on the relative importance of the 2 AH currents in 
pure systems (intrinsic regime), and on the role of extrinsic effects (impurities).
Here, we show that the AH effect in a high-purity ferromagnet MnSi reveals a remarkable constancy.  At 
temperatures $T<T_C$, the AH conductivity $\sigma_{xy}^A$ is strictly proportional to $M$
with a proportionality constant $S_H$ that is independent of both $T$ and magnetic field $\bf H$.

Conventionally, the observed Hall resistivity $\rho_{yx}$ 
in a ferromagnet is written empirically as~\cite{review}
\be
\rho_{yx} = R_0B + \mu_0R_sM
\label{rxy0}
\ee
where $R_0$ is the ordinary Hall coefficient, $\mu_0$ the 
permeability and ${\bf B} = \mu_0({\bf H + M})$ the induction field.  The
``anomalous Hall coefficient'' $R_s(T)$ is a scale factor 
that matches the $M$--$H$ curve to the anomalous part of the Hall resistivity
$\rho_{yx}' \equiv \rho_{yx}-R_0B$.
As such, $R_s(T)$ must be independent of the field $B$.  AH measurements are routinely
reported as a plot of $R_s(T)$ vs. $T$ as an empirical parameter.
Yet, Eq. \ref{rxy0} has never been justified microscopically.


To distinguish between the two theories, we have focused on the intrinsic
AH signal found in high-purity ferromagnets.  In these systems, with large 
magnetoresistance (MR), the difficulties with Eq. \ref{rxy0} become acute.
Additivity of currents in a solid implies that the total 
Hall conductivity is the sum
$\sigma_{xy} = \sigma_{xy}^N + \sigma_{xy}^{A}$
where $\sigma_{xy}^N$ is the ordinary Hall conductivity.
Additivity also requires that $\sigma_{xy}^A$ be proportional 
to $M$~\cite{Zeng06,Onose,review, Aeppli,Anke06}, which we express as 
\be
\sigma_{xy}^A = S_HM.
\label{sxy}
\ee
The scale factor $S_H$ plays the central role in our analysis.  
Converting $\sigma_{xy}$ to $\rho_{yx}$, we have 
\be
\rho_{yx} = R_0B + S_H\rho^2M
\label{rhoxy}
\ee
with $\rho$ the resistivity and $R_0 = \sigma_{xy}^N\rho^2/B$ 
(we assume that $\rho_{yx}\ll\rho$).  
We remark that Eq. \ref{rhoxy} goes beyond just taking
$R_s$ to be $\rho$ dependent (e.g. see Ref. \cite{Kats04}).
When $\rho$ varies strongly with $H$, $M$ fails to match $\rho_{yx}'$ altogether, 
and $R_s$ cannot be either defined or measured.  This is especially so 
when $\ell$ changes greatly with $T$ and $H$.  As we show, focusing on $S_H$ uncovers 
the proper scaling between $M$ and the AH response.

\bfig
\incl[width=8.5cm]{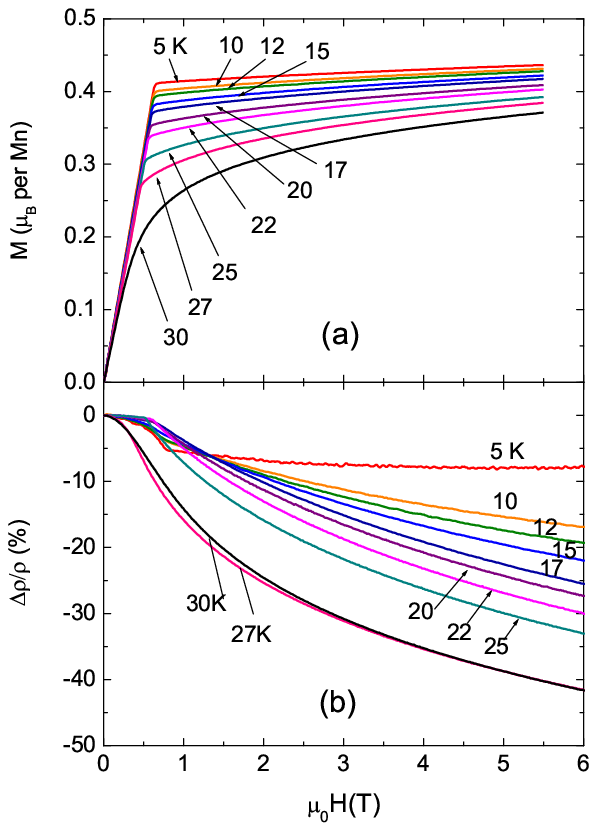}
\caption{\label{MvsH} 
(Color online)
 The magnetization curves $M$ vs. $H$ (Panel a) and 
curves of the relative magnetoresistance $\Delta\rho/\rho$ (Panel b) in MnSi at selected $T\le T_C$ (30 K).
Below $T_C$ in Panel a, the linear increase in $M$ up to the kink field $H_k$ reflects
the canting of the moments towards $\bf H$.  
In Panel b, the MR becomes large (40$\%$)
as $T\rightarrow T_C$.  Above $H_k$, $M$ displays a negative curvature
(especially near $T_C$) whereas the curvature in $\Delta\rho/\rho$ is positive. 
}
\efig
%
The metal MnSi, which displays one of the
highest conductivities in a ferromagnet, 
has drawn intense interest because it exhibits non-Fermi liquid behavior at applied pressures 
above 14 kbar~\cite{Pfleiderer97,Pfleiderer01}. 
Under hydrostatic pressure, it turns out that the Hall effect is indeed highly sensitive to the 
helical spin configuration, as discussed elsewhere~\cite {LeeOnose}.
At ambient pressure and zero $H$, MnSi undergoes a transition at $T_C$ = 30 K to a helical magnetic state
with a long pitch $\lambda$ ($\sim$180 \AA)~\cite{Ishikawa85}.  Neutron scattering experiments 
have established that, in a weak field ($H < 0.1$ T), the spins cant towards the direction of $\bf H$ 
to assume a conical structure and eventually  align at $H \sim 0.6$T ~\cite{Ishikawa84,Thessieu97}.   

The MnSi crystals were grown by the floating-zone method.  Three crystals 
of area 2$\times$1 mm$^2$ and thickness 50-80 $\mu$m were measured.   
At 4 K, values of $\rho$ range from 0.4 to 5 $\mu\Omega$cm.  The residual resistivity
ratio varies from 40 to 80.  Contacts with contact resistance $\le 1\,\Omega$ were made 
with Ag epoxy.   Hall measurements were performed with with the current (5 mA) 
applied parallel to the longest side ($x$ axis), the field $\bf H$ parallel to the shortest 
side ($z$ axis) and Hall $E$ field measured along $\bf \hat{y}$.
With field-sweep rates 0.05-0.1 T/min, we can resolve 
changes of $\sim$2 n$\Omega$ cm in $\rho_{yx}$ at low $T$.  $M$ is measured in a SQUID 
magnetometer.  Above 2 K, hysteretic behavior is not observed.

As shown in the magnetization curves (Fig. \ref{MvsH}a), the conical angle 
rapidly closes with field to produce the initial linear increase in $M$.  
The kink at $H_k\sim$0.6 T corresponds to alignment of the moments along $\bf H$.

Between $T_C$ and 4 K, $\rho$ in zero $H$ decreases by a factor of 10.  As shown
below, this is entirely due to an increase in $\ell$ (which reaches $\sim$ 240 \AA~at 4 K).   
Figure \ref{MvsH}b shows that the MR is large ($\sim 40\%$ near $T_C$, decreasing 
to $17\%$ at 10 K at 6 T).  The large changes in $\rho$ with $T$ and $H$ make MnSi ideal for 
investigating how the intrinsic AH signal changes with carrier scattering time.

\bfig
\incl[width=8.5cm]{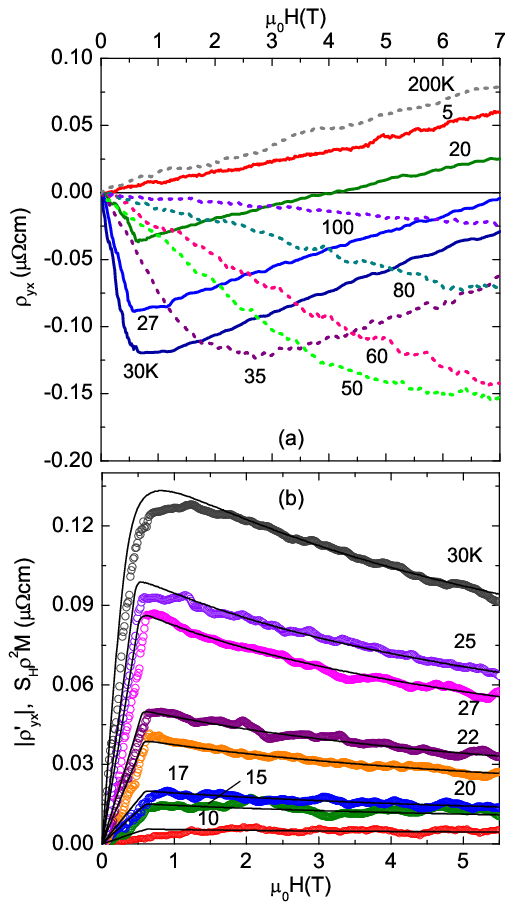}
\caption{\label{rxy} 
(Color online) 
The \emph{observed} Hall resistivity $\rho_{yx}$ in MnSi at selected $T$ (5--200 K, Panel a)
and the fit of the anomalous part $\rho_{yx}'$ to the modified magnetization profile
$\rho^2M$ below $T_C$ (Panel b).  Panel a shows that $\rho_{yx}$ is linear in $H$ at high $T$,
but gradually acquires an anomalous component $\rho'_{yx} = \rho_{yx}-R_0B$ 
with a prominent ``knee'' feature below $T_C$.    In Panel b, at each $T$, $\rho'_{yx}$ (open circles)
is fitted to the profile of $\rho^2M$ (solid curves), treating $S_H$ and $R_0$ as adjustable
$H$-independent parameters.  Note the positive curvature of the high-field segments.
}
\efig
%
Curves of $\rho_{yx}$ vs $H$ are shown in Fig. \ref{rxy} for $T$ from 5 to 200 K.
Above 150 K, $\rho_{yx}$ is linear in $H$, consistent with hole-like carriers ($\sigma_{xy}>0$ and thus $\rho_{yx}>0$). 
As $T$ decreases below 50 K, however, $\rho_{yx}$ develops strong curvature in weak $H$.  Below $T_C$,
$\rho_{yx}$ acquires an AH term that -- at first glance -- seems to resemble $M$ 
in accordance with Eq. \ref{rxy0}.
However, a more direct comparison reveals that the $M$--$H$ curves cannot be 
scaled to fit the curves of $\rho_{yx}'$.  The reason is their opposite curvatures.  
The curvature of $M$ is negative above $H_k$, whereas $\rho'_{yx}$ displays 
positive curvature (Fig. \ref{rxy}b).  The sign difference 
in the curvatures precludes definitively any satisfactory fit to Eq. \ref{rxy0}.

Our approach is as follows.  If Eq. \ref{rhoxy} is correct, at each $T<T_C$, 
the profile of $\rho_{yx}'$ vs. $H$ must match that of $\rho^2M$ 
vs. $H$ ($S_H$ is taken to be $H$-independent).  In this regard, it is reassuring that, unlike $M$, 
the product $\rho^2M$ shares the same curvature as $\rho_{yx}'$.
By varying the 2 parameters $R_0$ and $S_H$, we succeed in obtaining close fits at each $T$,
as shown in Fig. \ref{rxy}b.  The close match at each $T$ provides strong 
evidence for the validity of Eq. \ref{rhoxy}.  With $R_0$ and $S_H$ determined, 
the 2 Hall conductivities may be separated at each $T$
(see below).  

\bfig
\incl[width=8.5cm]{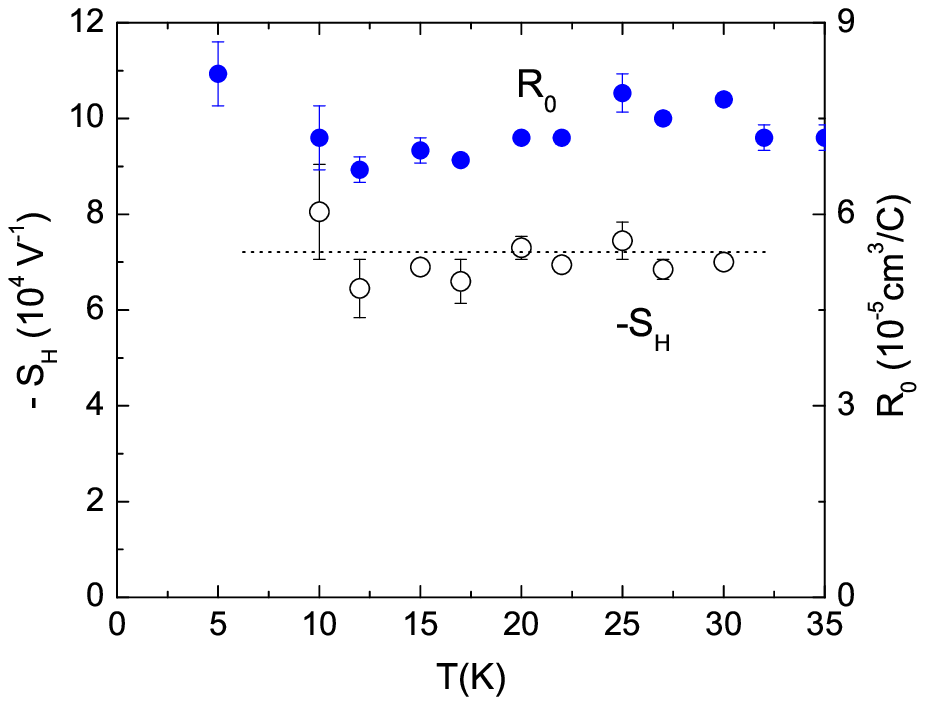}
\caption{\label{SH} 
(Color online) Values of the scale factor $S_H$ (open circles) and the ordinary Hall coefficient
$R_0$ (solid circles) obtained from the fits shown in Fig. \ref{rxy}b.  The average
of $S_H$ (dashed line) has the value $-(7.06\pm 0.48)\times 10^4$ V$^{-1}$.  The average
$R_0$ gives a Hall density $n_H =  (8.53\pm 0.07)\times 10^{22}$ cm$^{-3}$ 
and $k_F\simeq 1.36\times 10^8$ cm$^{-1}$ in Sommerfeld approximation.
}
\efig
%

We have also searched for a skew-scattering contribution by adding an AH conductivity that
scales as $M$ and is linear in $\ell$.  We write $\sigma_{xy}^{sk} = \alpha S_HM\rho(0)/\rho(H)$,
where the dimensionless parameter $\alpha(T)$ defines its magnitude at $H$ = 0 relative to the KL term,
and $\rho(0)/\rho(H)$ gives the $H$ dependence of $\ell$.
In Eq. \ref{rhoxy}, the second term is amended to $S_HM\rho(H)^2[1+\alpha\rho(0)/\rho(H)]$.  
We found that including $\alpha$ did not improve the fits.  Optimization leads to 
values of $\alpha$ that fluctuate from 0 to 0.05 with no discernible trend (and consistent with
$\alpha$ = 0).

As a consistency check, we note that 
the fits are physically meaningful only if both parameters
turn out to vary only weakly with $T$, if at all.  The variation of $R_0$ and $S_H$ obtained from
the fits are plotted against $T$ in Fig. \ref{SH}.  Within the scatter, the 2 parameters are virtually
independent of $T$.  The inferred $R_0$ is nearly unchanged 
as $T$ decreases from $T_C$ to 5 K, despite the
10-fold decrease in $\rho$.  
This verifies our starting assumption that the decrease in $\rho$
comes entirely from the increase in $\ell$, possibly reflecting suppression of scattering from
spin fluctuations.  From the Fermi wavevector $k_F\sim 1.36\times 10^8$ cm$^{-1}$, we find that 
the parameter $k_F\ell$ varies from 330 to 30 between 4 K and $T_C$.

The constancy of $S_H$ in Fig. \ref{SH} is more interesting and significant.
As $T$ decreases below $T_C$, both the magnetization and resistivity vary strongly with both
$H$ and $T$.  Nonetheless, the AH conductivity is completely
determined by $M(T,H)$, as expressed in Eq. \ref{sxy}.  The constancy of $S_H$
implies that the dependence of $\sigma_{xy}^A$ on $T$ (or $H$) derives entirely from 
that of $M(T,H)$.  In particular, the 10-fold change in $\ell$ below $T_C$
has no observable effect on $\sigma_{xy}^A$.
The task of predicting the AH conductivity in MnSi has been reduced to calculating
one constant, $S_H$.  This situation is in marked contrast to that presented by
an analysis based on $R_s(T)$ (Eq. \ref{rxy0}).

\bfig
\incl[width=8.5cm]{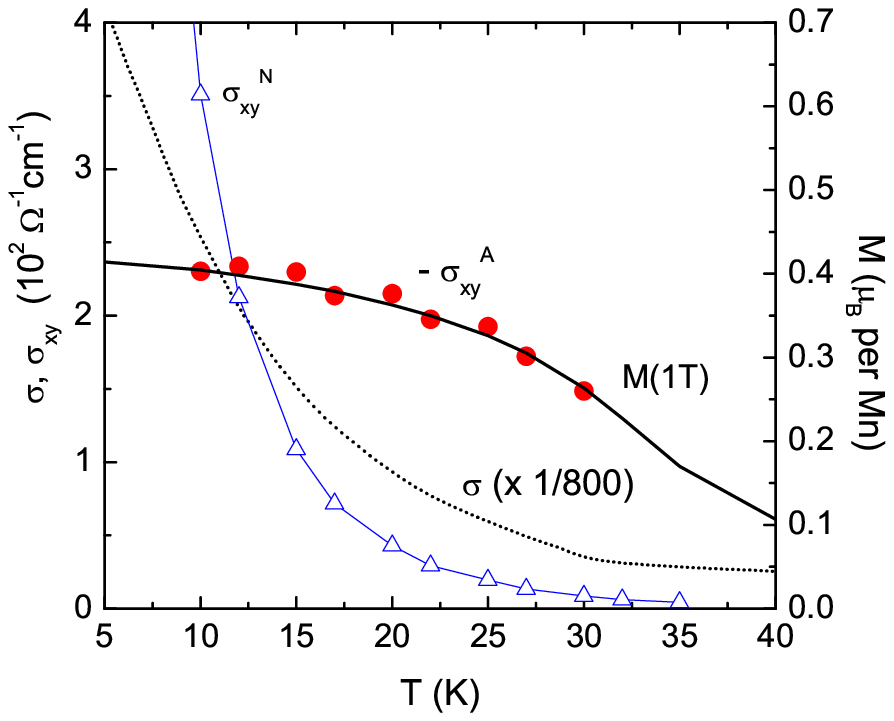}
\caption{\label{sigma} 
(Color online) Comparison of the AH conductivity $\sigma_{xy}^A$ (solid circles)
with the ordinary Hall conductivity $\sigma_{xy}^N$ (open triangles) in a 1-Tesla 
field (they have opposite signs).  $\sigma_{xy}^A$ is obtained from Eq. \ref{sxy} using 
$S_H$ shown in Fig. \ref{SH} and the measured $M$ (solid curve),
whereas $\sigma_{xy}^N\sim\ell^2$ is calculated from $R_0$.  As shown, $\sigma_{xy}^A$ 
is strictly independent of $\ell$; it changes slowly with $T$ only because $M(T)$ does.  
The $T$ dependence of the conductivity $\sigma\sim\ell$ (dashed curve)
reflects $\ell$ vs. $T$.  
}
\efig
%

It is instructive to compare directly the anomalous and ordinary Hall conductivities (Fig. \ref{sigma}).
As the latter (calculated as $\sigma_{xy}^{N} = R_0B/\rho(H)^2$) increases as $\sim\ell^2H$,
it greatly exceeds the former in magnitude at low $T$.  The two Hall conductivities are 
plotted in Fig. \ref{sigma} with the field fixed at 1 T.  As mentioned, below $T_C$, 
$\sigma_{xy}^A$ strictly follows the $T$ dependence of $M$ (solid curve), and is
insensitive to the steep increase in $\ell$ (the dashed curve shows the conductivity $\sigma$).  
At 5 K,  $\sigma_{xy}^A$ attains the value 240 ($\Omega$cm)$^{-1}$.  By contrast, $\sigma_{xy}^N$ 
is initially 20 times weaker than the AH term at $T_C$, but increases a 100-fold 
as $T\rightarrow$ 5 K and thus,  $\rho_{yx}'$ was not able to be detected in  Fig.\ref{rxy}.

Our finding that $\sigma_{xy}^A$ is nearly $T$ independent disagrees with Ref.~\cite{Aeppli}
which reports a strong deviation towards zero as $T$ decreases to 5 K.  
Our conjecture for the discrepancy is that the quantity plotted in Fig. 5 of Ref. \cite{Aeppli}
is actually the absolute value of the total Hall conductivity $|\sigma_{xy}| = |\frac{\rho_{yx}}{\rho^2}|$ (at a fixed field $H=$ 0.1T), rather than
$\sigma_{xy}^A$.   Between 30 and 5 K, $\rho_{yx}$ measured at 0.1 T falls towards 
zero as $T\rightarrow$ 5 K (see Fig. \ref {rxy} here), which implies that the total $\sigma_{xy}$ does the same.  It seems crucial to separate out the ordinary Hall current in MnSi.  
Another speculation for the discrepancy is that   at the low field  of $H = 0.1$ T  as in Ref.[14], the contribution of the magnetization to AHE in MnSi  may be non-trivial due to its helical nature.  Thus, its  contribution to $\rho_{yx}$ can be different from that of fully ferromagnetic state in $H>0.6$ T, as we observed under hydrostatic pressure \cite {LeeOnose}.

The relationship between the KL term 
and the Berry phase has been discussed by several groups~\cite{Niu,Nagaosa,JungwirthYao,Haldane}.  
The curl of the Berry potential leads to an effective magnetic 
field $\bf \Omega(k)$ in $\bf k$ space that adds a new term 
to the group velocity, viz. (see Ref. \cite{Ong} for an elementary treatment)
\be
\hbar{\bf v(k)} = \nabla_{\bf k}\epsilon({\bf k}) + e{\bf \Omega(k)\times E}.
\label{vk}
\ee
The anomalous term $e\bf \Omega\times E$, which is transverse to $\bf E$,
then gives a Hall conductivity that is independent of $\ell$ ({\it i.e.} dissipationless).

The results in Fig. \ref{sigma} showing that $\sigma_{xy}^A$ is insensitive
to the 10-fold change in $\ell$ from 5 K to $T_C$ provide
compelling evidence in favor of the KL theory (and its Berry phase-based versions).  
However, present theories do not account for the broad interval of $T$ over
which $\sigma_{xy}^A$ remains $\ell$-independent.  How ubiquitous the constancy
is (at low $T$) in other ferromagnetic systems \cite {Anke06} and how it is modified at higher 
$T$ are issues for further investigation (for e.g., in some ferromagnets, $\rho'_{yx}$ changes sign 
near $T_C$).  


The experiment also addresses the relative importance of 
skew scattering compared to the KL term~\cite{Nozieres,Bruno,Onoda06}.  
In recent calculations, the skew scattering term is either comparable
to the KL term~\cite{Bruno}, or strongly dominant when $k_{so}\ell > 1$ 
where $k_{so} = E_{so}/v_F$, $E_{so}$ and $v_F$ are the energy scale of spin-orbit 
interaction and Fermi velocity, respectively~\cite{Onoda06}.
When the skew term is included, there is apparently no regime in which the KL term is clearly dominant 
(i.e. $\sigma_{xy}^A$ strictly independent of $\ell$).

By contrast, our results on MnSi show that, at all $T<T_C$, 
$\sigma_{xy}^{sk}$ is essentially unresolved and $\sigma_{xy}^A$ is 
consistent with the KL term throughout the interval $30<k_F\ell<330$.  
In our experimental results, $\sigma^A_{xy}$ still remains constant 
at  $T < T_c$. This disagreement suggests either that skew scattering may 
have been greatly overestimated or that inelastic scattering may 
play a role at finite temperature\cite {Onoda}.

Further understanding of the intrinsic AH conductivity requires measurements in
high-purity crystals with large $k_F\ell$ (in thin-film samples
extrinsic scattering from the surface is problematic).
We show that additivity of Hall currents provides the correct perspective
to reconcile the large MR with the scaling between the AH current with $M$.  The new analysis
allows $R_0$ (hence $\sigma_{xy}^N$) to be isolated.  More significantly, it uncovers a scaling 
factor $S_H$ that is independent of both $H$ and $T$ below $T_C$.  The 
skew scattering contribution is negligibly small (0--5 $\%$).  
The constancy of $S_H$ implies the AH current is completely determined by the curves 
of $M$ vs. $H$ below $T_C$.  This simple scaling is obscured if $\rho_{yx}'$
is forced to fit $M$ in order to extract $R_s$, or if samples with large extrinsic scattering
are used.  

We have benefitted from useful discussions with N. Nagaosa and S. Onoda.  
Research at Princeton University was supported by the U.S.
National Science Foundation (DMR 0213706).

%

\begin{thebibliography}{99}

\bibitem{Karplus}  R. Karplus and J. M. Luttinger,   
 Phys. Rev.  {\bf 95}, 1154-1160 (1954).

\bibitem{Niu} G. Sundaram and Q. Niu, 
Phys. Rev. B {\bf 59}, 14915-14925 (1999).

\bibitem{Nagaosa}  M. Onoda and N. Nagaosa,
 J. Phys. Soc. Jpn. {\bf 71}, 19 (2002).

\bibitem{JungwirthYao} T. Jungwirth, Q. Niu and A. H.  MacDonald, 
 Phys. Rev. Lett. {\bf 88}, 207208 (2002); Y. Yao,  \etal., Phys. Rev. Lett.  {\bf 92}, 037204 (2004).


\bibitem{Haldane}  F. D. M. Haldane, 
Phys. Rev. Lett.  {\bf 93}, 206602 (2004).

\bibitem{Smit}  J. Smit, 
Physica  {\bf 21}, 877 (1955).

\bibitem{Matl} P. Matl, N. P. Ong, Y. F. Yan, Y.Q. Li, D. Studbebaker, T. Baum and G. Doubinia,
  Phys. Rev. B  {\bf 57}, 10248-1025 (1998).

\bibitem{Tokura}  Y. Taguchi, Y. Oohara, H. Yoshizawa, N. Nagaosa and Y. Tokura,
 Science  {\bf 291}, 2573-2576 (2001).


\bibitem{WLee} W.-L. Lee, S. Watauchi, V. L. Miller, R. J. Cava and N. P. Ong, 
 Science {\bf 303}, 1647-1649 (2004).


\bibitem{Mathieu04}  R. Mathieu, A. Asamitsu, H. Yamada, K. S. Takahashi, M. Kawasaki, Z. Fang, N. Nagaosa, Y. Tokura,
 Phys. Rev. Lett.  {\bf 93}, 016602 (2004).


\bibitem{Zeng06}  C. Zeng, Y. Yao, Q. Niu, and H. H. Weitering, 
  Phys. Rev. Lett.  {\bf 96}, 037204 (2004).

\bibitem{Onose} Y. Onose and Y. Tokura, 
 Phys. Rev. B {\bf 73}, 174421 (2006).

\bibitem{review} Hurd, C., \emph{The Hall Effect in Metals and Alloys}, 
153 -- 182 Ch.5  (Plenum, New York 1972).  

\bibitem{Aeppli}  N. Manyala,  Y. Sidis, J. F. Ditusa, G. Aeppli, D. P. Young and Z. Fisk,
 Nature Materials {\bf 3} 255-262 (2004).

\bibitem{Anke06}   A. Husmann and L. J. Singh, 
 Phys. Rev. B {\bf 73} 172417 (2006).

\bibitem{Kats04}  Y. Kats, I. Genish, L. Klein, J. W. Reiner and M. R. Beasley, 
 Phys. Rev. B {\bf70}, 180407 (2004);J. K$\ddot{\rm o}$tzler and Woosik Gil, {\it ibid}. {\bf 72}, 060412(R) (2005)


\bibitem{Pfleiderer97}   C. Pfleiderer, G. J.  McMullan, S. R. Julian and G. G. Lonzarich, 
Phys. Rev. B {\bf 55}, 8330-8338 (1998).

\bibitem{Pfleiderer01}  C. Pfleiderer,  S. R. Julien and G. G. Lonzarich,   
Nature  {\bf 414}, 427- 429 (2001).

\bibitem{LeeOnose} Minhyea Lee,  W. Kang, Y. Onose and N. P. Ong, (to be published).

\bibitem{Ishikawa85}   Y. Ishikawa, Y. Noda, Y. J. Uemura, C. F. Majkzak  and G. Shirane,   
 Phys. Rev. B  {\bf 31}, 5884-5893 (1985).

\bibitem{Ishikawa84}   Y. Ishikawa and M. Arai,  
  J. Phys. Soc. Jpn. {\bf 53}, 2726-2733 (1984).

\bibitem{Thessieu97} C. Thessieu, C. Pfleiderer, A. N.  Stepanov and J. Flouquet, 
J. Phys. : Condensed Matter {\bf 9}, 6677-6687 (1997).

%
\bibitem{Ong} N. P Ong,  and  W.-L. Lee, 
cond-mat/0508236.

\bibitem{Nozieres}  P. Nozi\`{e}res, and  C. Lewiner,  
 J. Phys. (France) {\bf 34}, 901-915 (1973).

\bibitem{Bruno}  V. K. Dugaev, P. Bruno, M.  Taillefumier,  B. Canals,  and C.  Laroix, 
 Phys. Rev. B {\bf 71}, 224423 (2005).

\bibitem{Onoda06} S. Onoda, , N. Sugimoto, and N. Nagaosa,
 Phys. Rev. Lett. {\bf 97}, 126602 (2006).

\bibitem{Onoda} S. Onoda  and N. Nagaosa,  \emph{private communication}.

\end{thebibliography}
\end{document}